**Schottky photodetectors with transparent conductive oxides for photonic integrated circuits**

J. Gosciniak[1*], J. B. Khurgin[2#]

[1]*ENSEMBLE3 sp. z o.o., Wolczynska 133, 01-919 Warsaw, Poland*
*Corresponding authors: *jeckug10@yahoo.com.sg and #jakek@jhu.edu*
[2]*Electrical and Computer Engineering Department, Johns Hopkins University, Baltimore, Maryland 21218, USA*

**Abstract**
Silicon photonics has many attractive features but faces a major issue: inefficient and slow photodetection in the telecom range. New metal-semiconductor Schottky photodetectors based on intraband absorption address this problem, but their efficiency remains low. We suggest that by creating a junction between silicon and a transparent oxide with appropriate doping, which results in a real permittivity close to zero (known as the epsilon near zero or ENZ regime), detection efficiency could increase by more than tenfold. Using Aluminum Zinc Oxide (AZO) as an example, we design an optimized AZO/Si slot photonic waveguide detector that could potentially reach an efficiency of several tens of percent, in contrast to a few percent for a metal/Si Schottky detector. This increase is primarily due to the lower density of states in AZO compared to metal, along with superior coupling efficiency and strong absorption within a 10 nm slot.

**Introduction**
The rapidly growing demand of data traffic all around the world already redirects the industry to photonic integrated circuits (PICs) as an alternative to electronics that can be fabricated on traditional silicon wafers [**1, 2**]. In such a way, it takes advantage of the existing production infrastructure of the electronic semiconductor industry where the entire photonic systems with lasers, waveguides, filters, couplers, modulators, and photodetectors can be integrated into the chips [**3, 4**]. Among those different components, photodetectors and modulators are the main building blocks of the optoelectronic link that provide signal conversion between electronic and photonic domains [**3, 4**]. Compared to modulators, which perform electrical-to-optical signal conversion, photodetectors perform inverse conversion, i.e., from the optical to the electrical domain. In addition, integrating the detector on-chip can improve responsivity by reducing the volume of active material that can generate thermal noise [**5**].

Photonic Integrated Circuits (PICs) take full advantage of available electronic semiconductor technology that is based on CMOS-compatible materials [**4**]. Thus, photodetectors that are based on the same material platform are under a huge interest as they can provide low cost and high fabrication possibilities. However, as silicon is the preferred material of choice from this aspect, it cannot be used for the photodetection of radiation at the telecom band (1300-1600 nm) because the photon energy is lower than its energy bandgap (~1.12 eV).

For this reason, several detectors based on intraband absorption in Si-compatible materials have been suggested and put into practice. These detectors encompass photo-thermoelectric (PTE) detectors [**6-8**] employing 2D materials like graphene and bolometric detectors, which also make use of graphene [**9, 10**] or TCO materials [**11**]. These detectors exhibit a sub-picosecond temporal response due to their dependence on electron temperature rather than lattice temperature. The "hot" carrier temperatures can reach thousands of degrees and then decrease quickly, in the matter of hundreds of femtoseconds. However, it should be noted that the efficiency of PTE is relatively low [**7, 12**], while in bolometric detectors the large DC current is required to obtain reasonable AC signal [**9**]. PTE and bolometric detectors also suffer from excessive noise [**7, 9, 12, 13**].

A more promising photodetection approach that can address these limitations involves internal photoemission (IPE), where a Schottky barrier is established at a metal-semiconductor interface with an energy level lower (0.2-0.8 eV) than that of the Si bandgap [**14-20**]. Nonetheless, these detectors also exhibit low efficiency, primarily due to various factors. The most evident ones are: (a) the limited containment of optical modes within the metal, attributed to the high absolute permittivity of metals; (b) low transport efficiency caused by fast electron-electron (EE) scattering, preventing many "hot" carriers from reaching the barrier; and, perhaps most significantly, (c) the reflection of carriers that do manage to reach the barrier due to the high density of states near the Fermi level of the metal. The latter problem is partially alleviated in IPE detectors incorporating graphene[4], which points one in the direction of looking for alternative materials for IPE detection.

Here we propose substituting metal with a transparent conductive oxide (TCO) material [**21-32**], using aluminum zinc oxide (AZO) [**22, 23**] as an illustrative example. TCO materials have recently garnered attention in the context of nonlinear devices [**25-29**] and optical modulators [**30-32**], with a specific focus on their operation within the "Epsilon Near Zero" (ENZ) regime, where the real part of the permittivity approaches zero [**28-31**]. Notably, some remarkable results have been reported. Yet, there remains a stubborn fact: TCO materials respond to both electrical and all-optical modulation signals based on the plasma (Drude) dispersion of free carriers [**25, 26**]. Consequently, TCO devices inherently exhibit substantial insertion losses, thereby limiting their practical applications, despite a considerable number of enthusiastic publications. At the same time, in the photodetector large absorption is not an impediment, but rather an advantage, and it is perhaps here that TCO ability of absorb wide wavelength range of infrared and visible radiation in a small volume can make TCO's materials of choice for integrated Si-compatible photonics.

With this in mind, this paper presents an analysis of employing TCO for internal photoemission (IPE) detection. We highlight the advantages of TCO while also acknowledging potential drawbacks that could impact performance. Although our analysis is preliminary due to the scarcity of experimental data, it is worth noting that it suggests a potential improvement on the order of magnitude when compared to metal IPE detectors. We hope that this finding will capture the interest of the experimentalist community and contribute to the advancement of integrated silicon photonics

**Operation principle of the Schottky photodetector in metals and TCO's**

The operation principle of Schottky photodetectors relies on the absorption of photons by metals that is accomplished by internal photoemission (IPE) across the Schottky junction between metal (usually Au) and semiconductor (usually n-doped or p-doped Si) [**14-20**] and is illustrated in **Fig. 1a**. The energy $E$ is counted from the Fermi level; hence the bottom of conduction band is far below (at -5.5 eV to be precise). The carriers are photoexcited from within a slice of energies $-\hbar\omega<E<0$ and end up within the range $0<E<\hbar\omega$. The fraction of carriers with energies above Schottky barrier $\Phi$ has a chance of being ejected from metal into semiconductor. Note that the density of states in the range $-\hbar\omega<E<\hbar\omega$ does change much and this density is high since the Fermi wavevector is large - 12 nm$^{-1}$ [**33, 34**]. This density of states is much higher than wavevector in the semiconductor (Si) which is in the range of 0 to 0.3 nm$^{-1}$. This mismatch presents a serious obstacle to effective injection as will be explained further on.

In a typical TCO the situation is different as shown in **Fig. 1(b)**. The Fermi level is relatively small, about 0.65 eV for the AZO operating near ENZ point with carrier concentration of $N_c\sim7\cdot10^{26}$ m$^{-3}$ [**25-27**]. Since $\hbar\omega>E_F$ the carriers are excited from the filled states with $-E_F<E<0$ where the density of states is non-uniform. Therefore, excited states energy distribution $\hbar\omega-E_F<E<\hbar\omega$ maintains (until electron-electron (EE) scattering causes redistribution of "hot" carriers) the same parabolic distribution in the band with more carriers having energies above the barrier than in the case of metal with the same value of energy barrier. Furthermore, the Fermi wavevector is not as high as in metal, only 2.74 nm$^{-1}$ [**33. 34**], i.e., comparable to the Fermi wavevector in Si. All of it greatly improves the chances of electron injection into the Si.

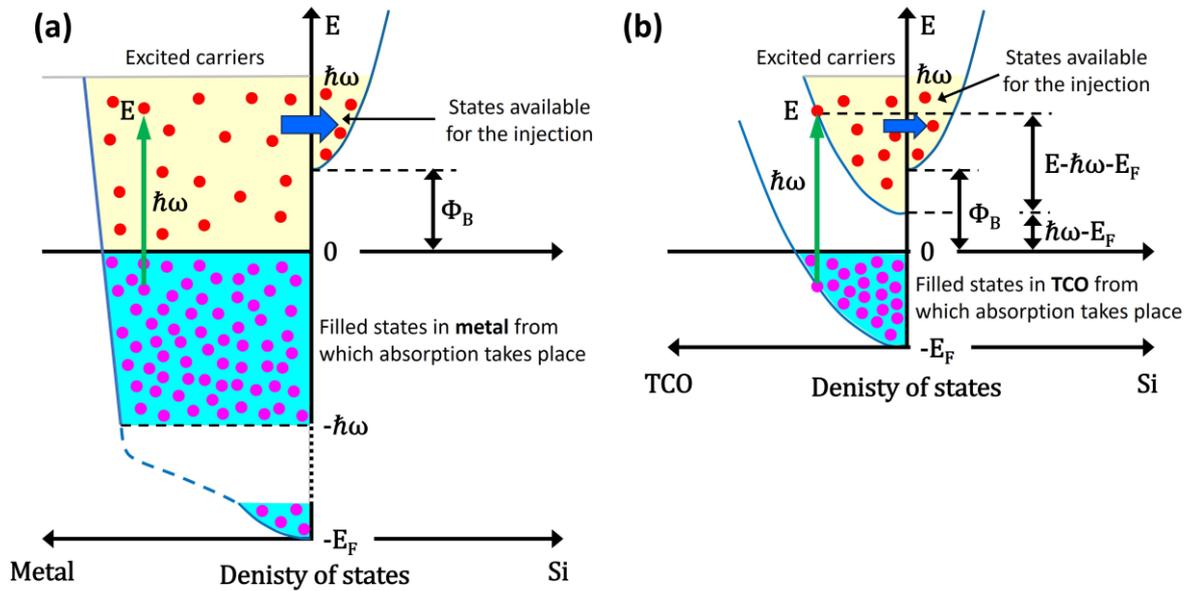

**Fig. 1**. Internal photoemission across the Schottky barrier between (a) metal and semiconductor and (b) TCO and semiconductor.

The process of photon-induced emission of electrons from metals to semiconductors is based on the Fowler proposal and was described by Spicer [21]. It consists of a three main steps: (1) generation of "hot" electrons in the metal through the absorption of light, (2) diffusion of a portion of "hot" electrons to the metal/semiconductor interface before thermalization, and (3) injection of "hot" electrons with sufficient energy and correct momentum into the conduction band of the semiconductor through the IPE process. We shall consider these steps one by one but first we will describe the layout of the proposed TCO detector.

**Proposed Schottky photodetector layout and field concentration in it.**
The proposed photodetector arrangement is shown in **Fig. 2a** and comprises a thin layer of TCO material with a thickness of 10 nm, placed between two 20nm thick layers of n-doped Si ($N_c = 5 \times 10^{24}\ m^{-3}$) and topped with a rib of undoped Si that 400 nm wide 260 nm high. The Si and AZO are separately contacted and reverse biased is applied to the Schottky junction formed on both top and bottom surfaces of TCO layer. The Schottky barrier height between TCO and silicon lays in a desired energy range of 0.4-0.6 eV [**35-40**] for detecting radiation of NIR wavelength of 1550 nm (0.8 eV). Different combinations of TCO's and semiconductors like SiGe or TiO₂ and etc., can be used to change the barrier height to adjust for operation at different wavelengths.
In previous Schottky photodetectors, plasmonic waveguides were utilized [**14-20**], but they suffered from poor coupling efficiency between photonic waveguides and photodetectors, resulting in reduced responsivity (as indicated in Ref. [**14**] and [**17**]). In our proposed waveguide-integrated photodetector, our simulations suggest that the coupling efficiency exceeds 90%, thereby increasing the amount of power contributing to photocurrent generation.

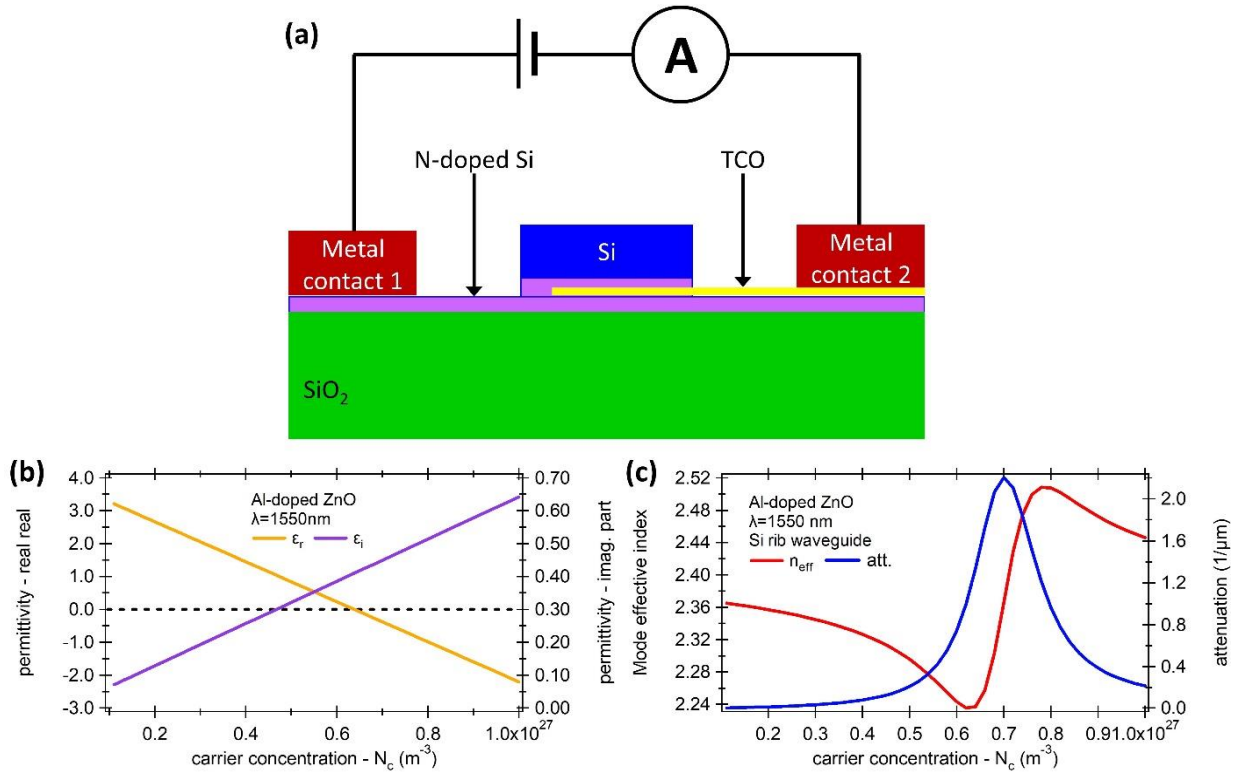

**Fig. 2**. (a) Geometry of the proposed TCO-based photodetector organized in the photonic rib waveguide arrangement. (b) Dispersion of complex permittivity (real and imaginary parts) of AZO as a function of wavelength. (c) Mode effective index and mode attenuation as a function of carrier concentration.

The proposed TCO detector offers many noteworthy advantages. The first is its impressive coupling efficiency. The second, equally significant feature, is its capacity to concentrate optical energy within an ultra-thin slot, just a few nanometers thick. In this confined space, the energy is efficiently absorbed close to the Schottky barriers, which facilitates emission across these barriers. This capability arises from the continuity of the normal component of electrical displacement at the interface of two materials. As seen from **Fig. 2b** the permittivity of the AZO changes as a function of Al doping and reaches minimum (ENZ point) at around $N_c \sim 7 \cdot 10^{26} \ m^{-3}$ which causes field to concentrate inside AZO layer with ensuing increase in absorption. The mode attenuation at ENZ point was calculated at $\alpha_{abs}$=2.2 µm$^{-1}$ (**Fig. 1c**) thus, less than 1 µm long photodetector is sufficient to absorb half of the power coupled to a photodetector.

The significant field enhancement at the ENZ point is visually evident in **Fig. 3**, where the field distribution in the TM mode is depicted in both the ridge waveguide without (on the left) and with AZO (on the right). This results in nearly a tenfold increase in the field strength, which corresponds to a remarkable 100-fold enhancement in energy density. The process of field enhancement in thin layer of TCO material arranged in photonics and plasmonic waveguide structure was in detail described in our previous papers in Ref. [25, 26]. It is worth highlighting that the real part of the effective index experiences a marginal change of just 0.2 %. This suggests that light can be readily coupled from the conventional Si rib waveguide into the detector.

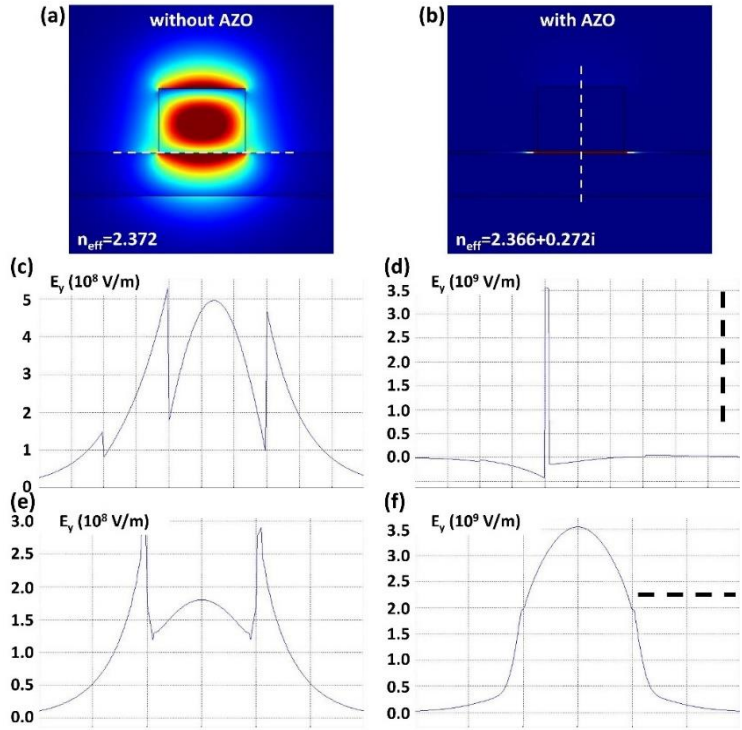

**Fig. 3**. (a, b) Electric field profile ($E_y$ component) for structure (a) without AZO layer and (b) with the AZO layer at the ENZ wavelength. (c, d) Electric field profile ($E_y$ component) in normal to the TCO layer (dashed white line in b) and (e, f) along the TCO layer (white dashed line in a) for structure (c, e) without TCO layer and (d, f) with TCO layer.

**"Hot" carrier generation mechanisms in TCO**

Having explored the "extrinsic" benefits of the suggested TCO Schottky photodetectors, which encompass efficient coupling and absorption in thin layers, our focus now shifts to the "intrinsic" characteristics that, in our view, are poised to greatly elevate the performance of the proposed TCO Schottky detector above that of metal-based ones.

We start by delineating the difference between the "hot" carrier generation mechanisms in metals and in TCO materials. According to Ref. [**41, 42**], four mechanisms exist but only two are relevant to the TCOs. Interband absorption is obviously not pertinent due to ZnO bandgap (3.44 eV) being much larger than photon energy of 0.8 eV. Furthermore, absorption assisted by electron-electron (EE) scattering occurs only for Umklapp processes, when one of the photoexcited "hot" carriers ends up in the adjacent Brillouin zone. In noble metals Fermi surface practically touches the Brillouin zone boundary making Umklapp processes highly probable, but in TCO's the Fermi surface is deep inside the Brillouin zone, therefore absorption enabled by EE scattering is not allowed. It is worth noting that while EE-assisted absorption is inhibited, the EE scattering, the primary process for carrier thermalization, is allowed and will be further discussed. Both interband and EE scattering assisted absorption produce lower energy carriers, hence their absence benefits efficiency of injection across the Schottky barrier as elaborated in the next section.

Two applicable absorption mechanisms remain. The first is absorption facilitated by phonons, impurities, and defects. In contrast to noble metals, all TCOs including AZO, are highly polar materials. The relevant phonon scattering mechanism in these materials is scattering by longitudinal optical (LO) phonons [**43, 44**] – a more potent process than the acoustic phonon scattering prevalent in metals. The scattering rate by LO phonons in ZnO was estimated using theoretical framework and found to be $\gamma_{LO} \sim 0.7 \cdot 10^{14}\ s^{-1}$ comparable to that in another polar wurtzite material GaN, on the scale of $\gamma_{LO} \sim 0.5 - 1 \cdot 10^{14}\ s^{-1}$ [**45**]. However, this scattering rate is considerably lower than the measured rate of up to $2.8 \cdot 10^{14} \, s^{-1}$ at $N_c = 6.67 \cdot 10^{26} \, m^{-3}$ obtained by fitting the experimental data at different electron densities in Al-doped ZnO [**27**]. Clearly, some other mechanism must be present, and that mechanism is scattering by ionized Al$^{3+}$ donors which becomes dominant at densities exceeding $\sim 10^{26} \, m^{-3}$. By applying the theory of scattering by screened ionized donors in degenerately doped semiconductors [**43**] we derive $\gamma_d \sim 3 \cdot 10^{14} \, s^{-1}$ at $N_c = 7 \cdot 10^{26} \, m^{-3}$. Therefore, scattering by ionized donors emerges as the dominant mechanism contributing to absorption. Unlike phonon and especially EE scattering, ionized donor scattering is elastic, with significant implications [**43, 44**].

The last mechanism is Landau damping or surface-assisted absorption, which is absorption enabled by the fact that when an electron collides with the surface (or the "wall"), momentum can be transferred between the electron and the entire lattice, similar to when an electron collides with a phonon or defect. The absorption rate due to Landau damping (LD) is expressed by $\gamma_{LD} = 3v_F/8d_{eff}$, where $v_F$ is the Fermi velocity and the effective thickness is defines as $d_{eff} = \int E^2 \, dV / \int E_\perp^2 \, dS$, i.e., the volume to surface ratio of the mode in the TCO and $E_\perp$ is the normal component of the electric field at the TCO-Si interface. For a considered structure not-surprisingly $d_{eff}$~5 nm, i.e., one half of the TCO thickness, while Fermi velocity is 0.9·10$^6$ m/s which brings LD rate to $\gamma_{LD}$=7·10$^{13}$ s$^{-1}$, or about a quarter of the ionized donor scattering rate $\gamma_{id}$. The "hot" carriers generated by surface collisions, i.e., Landau damping, are all located within a thin layer of thickness $\Delta L = v_F/v$~4.2 nm, from the surface where $v$ is optical frequency (200 THz for wavelength of 1550 nm).

In TCOs, all absorption processes, including phonon/ionized donor-assisted and LD, generate "hot" carriers with energy as high as $\hbar\omega$ above Fermi level. This presents an advantage for TCOs over metals, where EE-assisted scattering generates lower-energy "hot" carriers. However, to leverage this advantage, it is crucial for the "hot" carriers to reach the Schottky barrier without energy loss.

**Hot carrier transport efficiency**

Before reaching the Schottky barrier, "hot" carriers may undergo a scattering process, often referred to as a "collision," which can either preserve or reduce their energy. If the collision is purely elastic, such as in the case of ionized donor scattering, only the momentum of the "hot" carrier is affected, while their energy remains unchanged. Additionally, the direction of momentum is minimally altered, with a preference for forward scattering. However, the situation differs when LO phonon scattering comes into play. In this scenario, the energy of electrons is reduced by the energy of an LO phonon, which is approximately 0.072 eV [**46**]. This reduction is significant enough to have a detrimental impact on the likelihood of emission over the energy barrier. Nonetheless, the most significant factor is EE scattering. During each EE-scattering event, the energy of the "hot" electron is distributed among three entities: the original electron and a newly generated electron-hole pair, which usually makes the energy insufficient to overcome the barrier. It should be reiterated here, that while the EE-scattering assisted absorption (Umklapp process) is not allowed in TCO, the EE scattering that does not involve photon is indeed permissible.

In the absence of dependable experimental data for AZO, we conduct calculations for total (which encompasses spin-preserving and spin-flip) EE scattering, following the theoretical framework [**44**], and obtain the result $\gamma_{EE}$=6·10$^{13}$ s$^{-1}$ which is somewhat less that EE scattering rate in gold (10$^{14}$ s$^{-1}$) which may be attributed to the stronger screening by polar lattice in ZnO. We can now establish the mean free path length ($L_{MFP}$) for quasi-ballistic carriers, which refers to carriers that have undergone only elastic collisions as $L_{MFP} \sim v_{hot}/(\gamma_{LO} + \gamma_{EE})$, where $v_{hot}$ is the velocity of "hot" electron with energy above the barrier, i.e., on average about $v_{hot} \sim 0.95 \times 10^6 \, m/s$. Consequently, the mean free path for quasi-ballistic carriers is approximately 8 nm. Given the 10 nm thickness, this suggests that at least 70-80 % of "hot" carriers will successfully reach one of the interfaces. As a result, we can draw the conclusion that the transport efficiency in TCOs surpasses what can be attained in metals. This is primarily due to the fact that light is absorbed within a narrow barrier, and Schottky barriers are established on both sides of this barrier.

**"Hot" carriers' injection efficiency**

The quasi-ballistic carriers whose energy distribution is depicted in **Fig. 1a and b** arrive at the interface with angular distribution R(θ) - for carriers generated through phonon/impurity-assisted absorption their distribution is $R_{ph,i} = \frac{3}{4}cos^2\theta + \frac{1}{4}$, while for the carriers generated via Landau damping (LD) is $R_{LD}(\theta) = 2|cos^3\theta|$ [**41, 42**].

For TCO, about 75% of carriers arise from impurity scattering, and the remaining 25% are due to LD, leading to the effective distribution $R_{eff}(\theta) = \frac{1}{16}(8|cos^3\theta| + 9\,cos^2\theta + 3)$. For metals we assume a similar distribution.

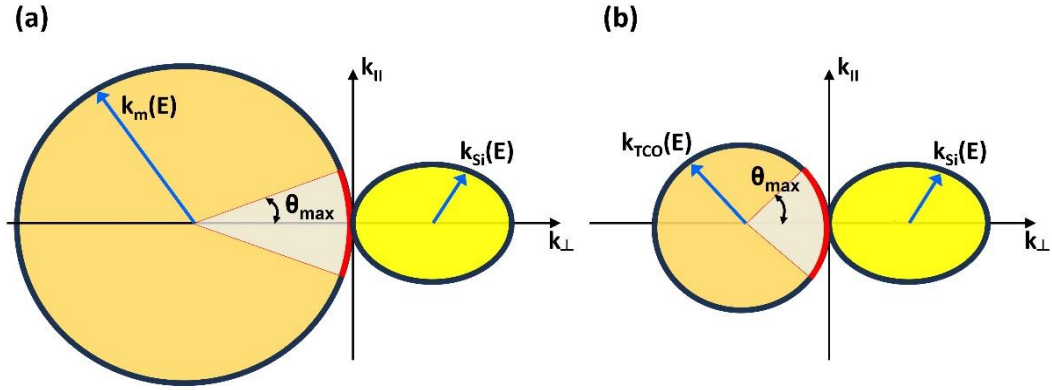

**Fig. 4.** Density of states and ranges of in-plane wavevectors $k_\parallel$ and corresponding angles $\theta_{max}$ that allow injection into Si from (a) metal (Au) and (b) TCO (AZO).

The electron wavevector in the "emitter", whether it's a metal or TCO, is equal to $k_{m(TCO)}(E) = \sqrt{2m_{m(TCO)}(E+E_F)/\hbar^2}$, where the effective mass of TCO (here Al-doped AZO) at the Fermi level of 0.65 eV is assumed at $m_{TCO} = 0.35m_0$ [**27, 33, 34**] as shown in **Fig. 4**. The states located on the spherical surface with this radius possess sufficient energy to surmount the barrier. However, due to the conservation of in-plane momentum, only states with an in-plane wavevector $k_\parallel$ smaller than wavevector in silicon, $k_{Si}(E) = \sqrt{2m_T(E-\Phi)/\hbar^2}$, where $m_T = 0.2m_0$ [**41, 42**] is a transverse effective mass in silicon. These states are graphically shown as a cone in **Fig. 4** with maximum permissible injection angle, subtending this cone:

$$\theta_{max}(E, \Phi_B) = \sin^{-1}\sqrt{\frac{m_T}{m_{m(TCO)}}\frac{E-\Phi_B}{E_F+E}} \quad (1)$$

The angular injection range for Transparent Conductive Oxides (TCO) is notably wider when contrasted with that of metallic materials. For example, assuming reasonable values of *E*=0.7 eV, i.e., slightly lower than a provided photon energy of 0.8 eV, $\Phi_B$=0.6 eV, $E_{F,TCO}$=0.65 eV and $E_{F,m}$=5.5 eV for TCO and gold respectively, one obtains θ$_{max,TCO}$=11.9° and θ$_{max,m}$=3.2° which, as anticipated, is poised to have substantial implications. Indeed, the extraction efficiency is:

$$\eta_{ext}(\hbar\omega, \Phi_B) = 2 \times 2 \int_{E_{0,m(TCO)}}^{\hbar\omega} f_{m(TCO)}(E,\hbar\omega) \int_0^{\theta_{max}(E,\Phi_B)} T(\theta)R_{eff}(\theta)\sin\theta\, d\theta dE \quad (2)$$

where the first factor of two arises from the degeneracy of two longitudinal valleys in Si into which injection takes place. The other factor of 2 stems from presence of the Schottky barrier on both sides of the absorbing layer (the metal detector with thin layer of metal in the middle of the plasmonic mode has been previously proposed in Ref. [19]. Here, it allows us a fairer comparison between TCO and metal). According to **Fig. 1a** in the case of metal, the lower limit of energy is $E_{0,m} = 0$, and the normalized energy distribution is uniform, $f_m = 1/\hbar\omega$. In contrast, **Fig. 1b** illustrates the situation with TCO, where the lower limit of energy is $E_{0,TCO} = \hbar\omega - E_F$, and the normalized energy distribution exhibits a parabolic profile, $f_{1,TCO}(E, \hbar\omega) = (3/2)\left(1/(2E_F - \hbar\omega)^{3/2}\right)(E + E_F - \hbar\omega)^{1/2}$. Subsequently, the transmission coefficient is determined as

$$T(\theta) = 1 - \frac{\frac{k_{m(TCO)\perp}}{m_{m(TCO)}} - \frac{k_{s,\perp}}{m_L}}{\frac{k_{m(TCO),\perp}}{m_{m(TCO)}} + \frac{k_{s,\perp}}{m_L}} \quad (3)$$

where the longitudinal components of wavevectors are $k_{m(TCO),\perp} = k_{m(TCO)}(E)\cos\theta$ and $k_{Si,\perp} = \sqrt{2m_L(E - \Phi)/\hbar^2 - k_{m(TCO)}^2 \sin^2\theta\, m_L/m_T}$. Prior to performing exact calculation, one can simplify the expression for extraction efficiency under assumption of all angles being small, i.e., $\sin\theta \approx \theta$ and obtain for the metal:

$$\eta_{ext,m}(\hbar\omega, \Phi_B) \approx 4\frac{R_{eff}(0)}{4\hbar\omega}\frac{m_T}{m_0}\frac{(\hbar\omega - \Phi_B)^2}{(E_F + \Phi_B)} \quad (4)$$

i.e., essentially a Fowler formula. For the TCO we additionally make a reasonable assumption that since the energy of carriers capable of surpassing the barrier lays in the range $\Phi_B < E < \hbar\omega$ we use approximation $E \sim (\hbar\omega + \Phi_B)/2$ for all the terms with exception of $E - \Phi_B$ so that

$$\eta_{ext,TCO}(\hbar\omega, \Phi_B) \approx \frac{3}{2}\frac{R_{eff}(0)}{E_F^{3/2}}\frac{m_T}{m_{TCO}}\frac{\left(E_F - \frac{\hbar\omega}{2} + \frac{\Phi_B}{2}\right)^{\frac{1}{2}}}{E_F + \frac{\hbar\omega}{2} + \frac{\Phi_B}{2}}(\hbar\omega - \Phi_B)^2 \quad (5)$$

which also resembles the Folwler's formula The ratio of the extraction efficiencies is then

$$\frac{\eta_{ext,TCO}}{\eta_{ext,m}} = \frac{3}{2}\frac{m_0}{m_{TCO}}\frac{E_{F,m} + \Phi_B}{E_{F,TCO}}\frac{\hbar\omega}{E_{F,TCO}}\frac{\left(1 - \frac{\hbar\omega - \Phi_B}{2E_{F,TCO}}\right)^{1/2}}{1 + \frac{\hbar\omega + \Phi_B}{2E_{F,TCO}}} \quad (6)$$

Three factors appear to contribute the most to the enhanced extraction efficiency of TCO. In order of importance, they are:

1. Much lower value of Fermi energy in TCO (almost a factor of 10)
2. Lower effective mass in TCO (factor of 3)
3. More favorable parabolic energy distribution of photoexcited carriers in TCO (factor of 1.5)

The final ratio is approximately on the scale of 1/2, suggesting that one can anticipate an enhancement by a factor of approximately 20-25. In **Fig. 5**, we present the extraction efficiency for metal (Au) (**Fig. 5a**) and TCO (AZO) (**Fig. 5b**). The dashed lines labeled as (I) in both graphs correspond to the approximations provided in **Eq. 4 and 5**, respectively. On the other hand, curves labeled as (II) are derived from the exact formula defined by **Eq. 2**, assuming perfect transmission ($T=1$) with the cone of allowed angles $\theta \leq \theta_{max}$. Evidently, the approximations remain valid, and upon examining the curves, the ratio of extraction efficiencies for TCO/Si and Au/Si junctions falls within the range of 24 to 26, consistent with the estimate presented in **Eq. 6**. Upon accounting for the partial reflection at the interfaces, the extraction efficiency, as indicated by curves (III), experiences a reduction. For Au/Si interfaces, this reduction ranges from a factor of two at very low barrier heights $\Phi_B$=0.2 eV to a factor of 3.5 at $\Phi_B$=0.6 eV. In the case of AZO/Si interfaces, the reduction is somewhat less, and the ratio of extraction efficiencies falls within the range of 29-32, as illustrated in **Fig. 6**, curve (a). The additional increase in the ratio can be attributed to the fact that the velocity mismatch responsible for reflections in **Eq. 3** is lower for TCO/Si junctions compared to Au/Si junctions.

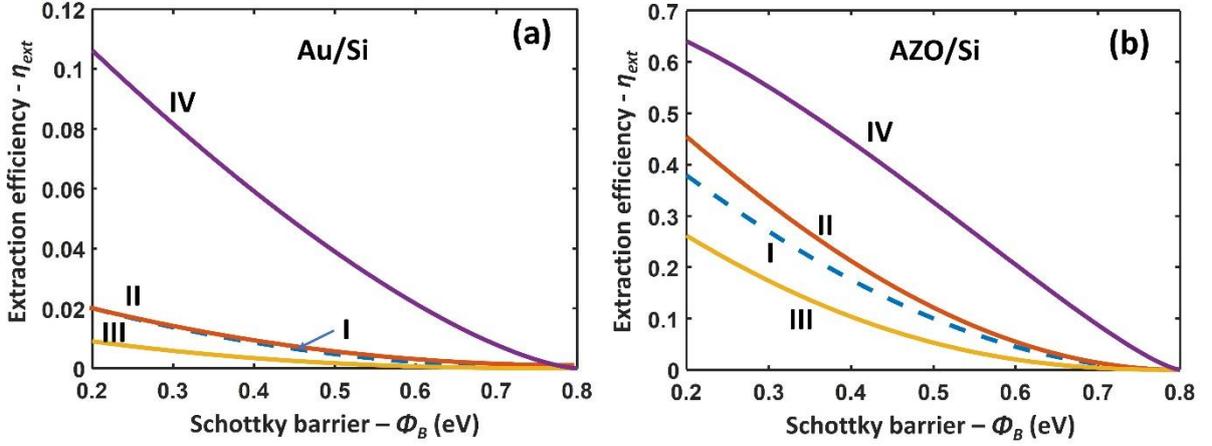

**Fig. 5.** The dependence of the extraction efficiency of the "hot" carriers on barrier height for (a) Au/Si and (b) AZO/Si junctions. The graphs in both (a) and (b) correspond to (I) – approximations **Eq. 4** and **Eq. 5**, (II) exact result for a smooth interface without considering reflections, (III) exact results for smooth interface, accounting for reflections as per **Eq. 2**, and (IV) – results for rough interface with complete relaxation of momentum conservation restrictions as in **Eq. 7**.

Numerous experimental [**47-49**] and theoretical [**49**] works have shown that the "hot" carriers extraction efficiency frequently surpasses the predictions established for smooth interfaces when roughness is introduced at the interface. This phenomenon is primarily attributed to the relaxation of the in-plane momentum conservation rule, particularly in the case of highly irregular interfaces where this rule is no longer strictly enforced. Consequently, electrons acquire the ability to transition into any state above the energy barrier within silicon. Additionally, they can undergo reflection into any state within the metal or TCO, provided that energy conservation is upheld. The permissible scattering states all lie upon the spherical surface of equal energy within the metal or TCO, or upon an ellipsoidal surface within silicon, as shown in **Fig. 4**. According to Fermi Golden Rule, the rate of scattering in a given direction depends only on the density of final states so, assuming densities of states $\rho_{m(TCO)}$ and $\rho_{Si}$, the extraction efficiency can be expressed as

$$\eta_{ext,m(TCO)}(\hbar\omega, \Phi_B) = \int_{\Phi_B}^{\hbar\omega} f_{m(TCO)}(E) \frac{2\rho_S(E)}{2\rho_S(E) + \rho_{m(TCO)}(E)} dE$$

$$= \left(\frac{m_s}{m_{m(TCO)}}\right)^{3/2} \int_{\Phi_B}^{\hbar\omega} f_{m(TCO)}(E) \frac{2\frac{(E-\Phi_B)^{1/2}}{(E+E_F)^{1/2}}}{2\frac{(E-\Phi_B)^{1/2}}{(E+E_F)^{1/2}}\left(\frac{m_s}{m_{m(TCO)}}\right)^{3/2} + 1} dE \qquad (7)$$

where the factor of two is due to extraction from both sides of the metal (TCO) layer, and the density-of-state effective mass for Si takes into account the fact that injection takes place only into two longitudinal valleys, $m_s = 2^{2/3} m_L^{1/2} m_T^{2/3} = 0.52 m_0$. The results are shown in **Fig. 5a, b** as curves (IV). The extraction coefficient for metals increases tenfold and reaches a respectable, if not spectacular, few percentage points. For the TCO, on the other hand the extraction efficiency can reach few tens of percent which puts them in the range of sensitivities normally associated with the best interband detectors. Given the superior operational speed of Schottky detectors, this represents a meaningful advancement. However, even further enhancement in the extraction efficiency is expected for a lower Fermi energy level that corresponds to the lower carrier concentration into TCO. In this scenario, the extraction efficiency of close to 80 % is expected.

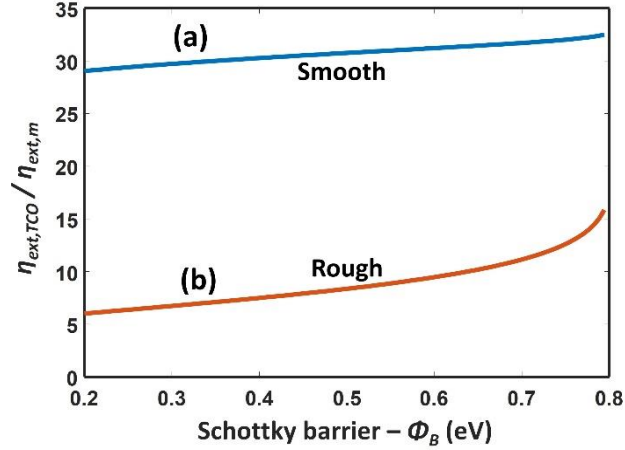

**Fig. 6**. Ratio of extraction efficiencies from AZO and Au. (a) Smooth interface and (b) rough interface, with the constraints of momentum conservation being entirely removed.

The enhancement in TCO extraction efficiency, compared to metals, varies from approximately 30-fold for smooth interfaces to less than 10 for rough interfaces. This behavior is anticipated, as the extraction efficiency at rough AZO/Si interfaces (curve IV in **Fig. 5b**) inevitably plateaus, unable to attain 100%. However, further reduction of the Fermi level in TCO below 0.65 eV, as examined in this paper, should bring the extraction efficiency closer to 100 %.

It is important to note that a meaningful comparison between metals and TCOs is only valid when the Schottky barrier heights and surface roughness are similar. While engineering these parameters is challenging, it is feasible. With this caveat, considering the substantial enhancement in extraction efficiency (as shown in **Fig. 5a and b**), paired with improved transport efficiency due to different scattering mechanisms, better coupling efficiency, and the ability to absorb light in a nanometer-scale layer, it is not unreasonable to anticipate a potential order of magnitude improvement in overall sensitivity by replacing metals with TCOs in Schottky interband detectors. Further enhancement in the extraction efficiency is expected for a thinner TCO layers as the distance to one of the interfaces can be reduced while the field enhancement in TCO will be further enhanced, thus, more energetic carriers can be produced with higher probability to reach one of the interfaces. The choice of 10 nm thickness of the TCO presented in this paper was solely motivated by the fabrication capabilities.

The novelty of this paper relies on replacing the metal by TCO material as an absorbing layer and two side transport of "hot" carrier through the interface in the Schottky-based photodetectors. Thus, over 60-fold for smooth interfaces and 20-fold for rough interfaces increases in the extraction efficiency can be expected what in combination with a high efficiency of light coupling to the photodetector can opens a new door in on-chip photodetectors.

**Conclusions**

In this study, we have investigated the potential use of transparent conductive oxides, specifically AZO, as a material for intraband Schottky barrier detection in the telecom range, with a focus on Si photonics. We have also provided a rough estimate of the detector's performance. While the proposed design here may not be considered optimal, and the obtained numerical values are all approximate, relying on several reasonable assumptions, the primary conclusion is unmistakable: the anticipated performance of the TCO-based detector is at least an order of magnitude superior to that of metal-based detectors. The factors contributing to the heightened sensitivity of the detector are as follows:

1. The utilization of a photonic waveguide in the Transparent Conductive Oxides (TCO) design, as opposed to a plasmonic waveguide, results in significantly improved light coupling efficiency into the detector when compared to metal-based detectors.
2. Absorption takes place entirely within the thin active TCO layer, in close proximity to Schottky barriers on both sides of the layer.
3. The predominant scattering mechanism within TCO materials involves scattering by ionized donors, which is an elastic process that does not alter the energy of "hot" carriers. This is in contrast to electron-electron scattering, which is prevalent in metals and leads to energy relaxation on a femtosecond time scale.
4. The energy distribution of photo-excited carriers in TCO materials exhibits is favorable skewed towards higher energies above the Schottky barrier, as compared to metals.
5. Of paramount importance is the fact that both the effective mass and Fermi wavevector, and thus the density of states, are substantially lower in TCOs when compared to metals. This characteristic significantly enhances the likelihood of electron injection into silicon.

With this, we express hope that this study will capture the attention of the scientific community and set in motion the experimental efforts required for the development of practical silicon-compatible detectors in the telecom wavelength range.

**Acknowledgements**

J.G. thanks the "ENSEMBLE3 - Centre of Excellence for Nanophotonics, advanced materials and novel crystal growth-based technologies" project (GA No. MAB/2020/14) carried out within the International Research Agendas program of the Foundation for Polish Science co-financed by the European Union under the European Regional Development Fund and the European Union's Horizon 2020 research and innovation program Teaming for Excellence (Grant Agreement No. 857543) for support of this work. J.K. has never been more obliged to his friends and colleagues Prof. P. Noir and Dr. S. Artois for their support and long stimulating discussions.